\documentstyle[12pt,epsfig]{article}

\begin{document} 

\title{Meson Effects on Chiral Condensate at Finite Density } 

\author{Mei Huang$^{1,2}$, Pengfei Zhuang$^{1}$, Weiqin Chao$^{2,3}$ \\ 
 {\small  $^1$ Physics Department, Tsinghua University, Beijing 100084, China}\\
 {\small  $^2$ CCAST, Beijing 100080, China}\\ 
   {\small $^3$ IHEP, Chinese Academy of Sciences, Beijing 100039, China   }}

\maketitle

\date{}

\begin{abstract}

Meson corrections on chiral condensate up to next-to-leading order in $1/N_c$ expansion
at finite density are investigated in NJL model with chiral symmetry explicitly breaking. 
Compared with the mean-field results,  the chiral phase transition is still of first order, 
while the properties at the critical density of chiral phase transition  are found to 
change significantly.

\end{abstract}

~~~PACS:  11.30.Rd, 11.10.Wx,12.38.Lg 

\newpage

\section{Introduction}

It is generally believed that QCD undergoes chiral restoration 
at high temperature and density.
The chiral condensate $<{\bar q}q>$, which is regarded 
as the chiral order parameter,
behaves quite differently as a function of 
baryon density $\rho$ as compared to its variation with temperature $T$.
At baryon density $\rho=0$, quark condensate $<{\bar q}q>$
does not change explicitly
till $T \simeq 0.9 T_c$, where $T_c$ is the critical temperature.
However, a model independent study \cite{rho}
shows that the quark condensate decreases
linearly with $\rho$ at low density. 
At normal nuclear matter density, $\rho=\rho_0=
0.17 fm^{-3}$, $<{\bar q}q>$ has already decreased to about $2/3$
of its vacuum value.

During the last decade, NJL model has been widely used to
investigate chiral phase transition at finite temperature and density. 
Most publications were based on mean-field approximation, 
i.e., quark self-energy in the leading order 
of $1/N_c$ expansion and meson in random-phase approximation (RPA). 
At this level, the excited meson modes
have no back interaction on quark self-energy.
A thermal system described by NJL model 
in the mean-field approximation behaves as an effective free fermion gas, i.e.,
only quarks contribute to the
thermal dynamical potential, while mesons do not play any role. 
This is clearly unphysical, since at least 
pion should play the dominant role at low temperature and
density \cite{heidberg}. 

Therefore, it is necessary to go beyond the mean-field approximation, i.e.,
considering meson corrections to quark self-energy.
In \cite{mesonT1} and \cite{mesonT2}, the meson effects on chiral
condensate at finite temperature have been investigated.
Our work in this paper  based on the self-consistent scheme 
of \cite{Ann} \cite{mei2} is to investigate the meson corrections on
the chiral phase transition at finite density with $T=0$.

\section{Meson Corrections at $T=0$ and $\mu=0$}

~
The two-flavor NJL model is defined through the Lagrangian density,
\begin{eqnarray}
\label{lagr}
{\cal L} = \bar{\psi}(i\gamma^{\mu}\partial_{\mu}-m_0)\psi + 
  G[(\bar{\psi}\psi)^2 + (\bar{\psi}i\gamma_5{\bf {\vec \tau}}\psi)^2 ],
\end{eqnarray}
here $G$ is the effective coupling constant of dimension ${\rm GeV}^{-2}$,
and $m_0$ is the current quark mass, assuming isospin degeneracy of the 
$u$ and $d$ quarks,
and $\psi, \bar{\psi}$ are quark fields with flavor, color and spinor
indices suppressed. 
  
It is not easy to give the full expressions of quark self-energy
and meson polarization functions self-consistently.
Usually an approximation scheme called large $N_c$ expansion 
is adopted.
V. Dmitra{\v s}inovi{\' c} et al. proved in \cite{Ann} 
that quark self-energy $\Sigma$
and meson's polarization function $\Pi_{M}$ 
shown in Fig. 1  are self-consistent to the subleading order
in $1/N_c$ expansion and can keep all the chiral relations
in the chiral limit.  
It is clear that the back interaction which conserves all the chiral
properties is reflected in the contributions from the 
meson propagator to the quark mass, namely in $\delta \Sigma$. 

Including current quark mass $m_0$, 
the gap equation for quark mass can be expressed as
\begin{eqnarray}
\label{gap}
m = m_0 + \Sigma_H +\delta \Sigma, 
\end{eqnarray}
where $\Sigma_H$ and $\delta \Sigma$ are the leading and subleading order of
quark self-enery in $1/N_c$ expansion, which can be
read directly from the Feynman diagrams,
\begin{eqnarray}
\Sigma_H=8iGN_cN_f m \int \frac{d^4p}{(2\pi)^4}\frac{1}{p^2-m^2},
\end{eqnarray}
\begin{eqnarray}
\label{self1}
& &\delta \Sigma=-8GN_cN_f m \int\frac{d^4pd^4q}{(2\pi)^8}
  [ \frac{(g^{(RPA)}_{\pi qq})^2}{q^2-m_{\pi}^2}
   (\frac{3}{(p^2-m^2)^2} \nonumber \\
& &   -\frac{3q^2}{(p^2-m^2)^2((p+q)^2-m^2)}  )
   +\frac{(g^{(RPA)}_{\sigma qq})^2}{q^2-m_{\sigma}^2}(
   \frac{1}{(p^2-m^2)^2}  \nonumber \\
& &  + \frac{2}{(p^2-m^2)((p+q)^2-m^2)}
   -\frac{q^2-4m^2}{(p^2-m^2)^2((p+q)^2-m^2)}) 
] ,
\end{eqnarray}
where $g^{(RPA)}_{\pi qq}$ and $g^{(RPA)}_{\sigma qq}$
are meson coupling constant in RPA.
At this position, it is necessary to point out that to keep the diagrams 
of meson corrections to quark self-energy and to meson polarization functions 
in the next-to-leading order in $1/N_c$ expansion properly, we have chosen
the internal meson 
propagators in pole approximation.

Evaluating one quark loop integral, one can get a simple relation between the 
quark condensate and  constituent quark mass $m$ \cite{Ann},
\begin{eqnarray}
-<{\bar q} q>=\frac{m-m_0}{4 G}.
\end{eqnarray}
 
Meson polarization function $\Pi_{M} ~(M=\pi, \sigma)$  can be written
as
\begin{eqnarray}
\Pi_{M}=\Pi_{M}^{(RPA)}+\delta \Pi_{M},
\end{eqnarray}
where $\Pi^{(RPA)}_{M} $ and  $\delta \Pi_{M} $ are pion
polarization functions in the leading and subleading order of $1/N_c$ expansion,
respectively,
and 
\begin{eqnarray}
\delta \Pi_{M} =\delta\Pi_{M}^{(b)} + \delta\Pi_{M}^{(c)} 
                    + \delta\Pi_{M}^{(d)} .
\end{eqnarray}

Using the standard way of calculating Feynman diagrams, 
it is easy to write down the RPA polarization functions,
\begin{eqnarray}
\Pi_{M}^{(RPA)} & = & 4iN_cN_f \int \frac{d^4p}{(2\pi)^4}
\frac{1}{p^2-m^2} \nonumber \\
& & - 2iN_cN_f(k^2-\epsilon_M 4 m^2)
\int \frac{d^4p}{(2\pi)^4}\frac{1}{(p^2-m^2)((p+k)^2-m^2)} 
\end{eqnarray}
with $\epsilon_{\pi}=0$ and $\epsilon_{\sigma}=1$.

\begin{figure}[ht]
\vspace*{-2truecm}
\centerline{\epsfxsize=16cm\epsffile{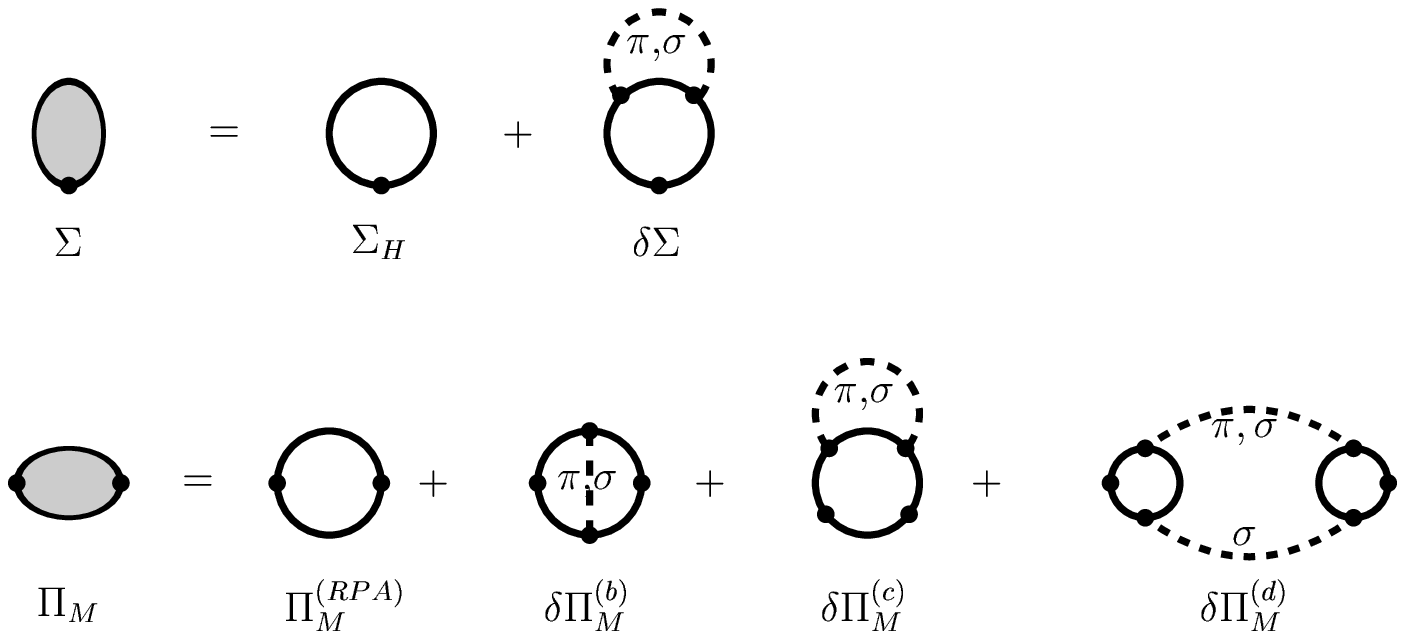}}
\vspace*{-10truecm}
\caption
{ Quark self-energy $\Sigma$ and meson polarization
function $\Pi_M$ in the quark and meson propagators. 
$\Sigma_H$ and $\delta \Sigma$ are the leading and subleading contributions to 
the quark mass. $\Pi^{(RPA)}_{M}$ and $\delta \Pi^{(b,c,d)}_{M}$ are 
the leading 
and subleading order contributions to meson polarization function. 
The heavy solid lines
indicate the constituent quark propagator,  
and the heavy dashed lines represent 
$\pi$ or $\sigma$ propagator $-{\rm i} D^{(RPA)}_{M}(q)$ in RPA
approximation. }
\label{kernel_fig}
\end{figure}

After calculating the trace of quark loops, the meson corrections to
the pion polarization function $\Pi_{\pi}$, corresponding to the three
Feynman diagrams in Fig.\ref{kernel_fig} can be expressed as:  
\begin{eqnarray}
\label{pib}
& & \delta\Pi_{\pi}^{(b)}(k) = 2N_cN_f\sum_{M=\pi, \sigma}\int\frac{d^4qd^4p}{(2\pi)^8}
\frac{ (g^{(RPA)}_{M qq})^2}{q^2-m^2_M} \nonumber \\
& & [\frac{1}{(p^2-m^2)((p+q-k)^2-m^2)}
+\frac{1}{((p+q)^2-m^2)((p-k)^2-m^2)} \nonumber \\
& &-\frac{k^2(q^2-\epsilon_M 4 m^2)}{(p^2-m^2)((p+q)^2-m^2)
((p-k)^2-m^2)((p+q-k)^2-m^2)}] ,
\end{eqnarray}

\begin{eqnarray}
& &  \delta\Pi_{\pi}^{(c)}(k) = -4N_cN_f\sum_{M=\pi,\sigma}
\int\frac{d^4qd^4p}{(2\pi)^8}\lambda_M\frac{ (g^{(RPA)}_{M qq})^2}{q^2-m^2_{M}} 
\nonumber \\
& &[\frac{1}{((p+q)^2-m^2)((p-k)^2-m^2)} + 
\frac{k^2(q^2-\epsilon_M 4 m^2)}{(p^2-m^2)^2((p+q)^2-m^2)((p-k)^2-m^2)} 
 \nonumber \\
& & +\frac{1}{(p^2-m^2)^2} + 
 \frac{2k\cdot q}{(p^2-m^2)((p+q)^2-m^2)((p-k)^2-m^2)} \nonumber \\
& &-\frac{k^2}{(p^2-m^2)^2((p-k)^2-m^2)} 
- \frac{(q^2-\epsilon_M 4 m^2)}{(p^2-m^2)^2((p+q)^2-m^2)}],
\end{eqnarray}

with the degeneracy $\lambda_\pi=3$, $\lambda_{\sigma}=1$ and
\begin{eqnarray}
& & \delta\Pi_{\pi}^{(d)}(k)=i\int\frac{d^4q}{(2\pi)^4}
\frac{ (g^{(RPA)}_{\pi qq})^2}{q^2-m^2_{\pi}}
\frac{ (g^{(RPA)}_{\sigma qq})^2}{(q-k)^2-m^2_{\sigma}} \nonumber \\
& & [\int\frac{d^4p}{(2\pi)^4}\frac{8mN_cN_f(k \cdot q-(p^2-m^2))}
{(p^2-m^2)((p+q)^2-m^2)((p+k)^2-m^2)}]^2.
\end{eqnarray}

In the similar way, 
we can get the meson corrections to the sigma polarization 
functions. However, it should be careful that 
the upper meson propagator for $\delta \Pi_{\pi}^{(d)}$
in Fig. 1 can only be $\pi$, but for  
$\delta \Pi_{\sigma}^{(d)}$ it can be $\pi$ and $\sigma$.

With the above quark self-energy and meson polarization functions,
 the meson mass $m_{M}$ is determined through pole condition 
\begin{eqnarray}
\label{pole}
1-2G\Pi_{M}(k^2=m_{M}^2)=0,
\end{eqnarray}
and the coupling constant $g_{M qq}$ is given by the residue at the
pole
\begin{eqnarray}
\label{couple}
g_{M qq}^{-2} = \partial \Pi_{M} /\partial k^2|_{k^2=m_{M}^2}.
\end{eqnarray}

Another important quantity is the pion decay constant $f_{\pi}$ which
is calculated 
from the vacuum to one-pion axial-vector matrix element. Replacing one
vertex 
$i \gamma_5{\bf{\vec \tau}}$ in $\Pi_{\pi}$ by 
$ig_{\pi qq} \gamma_5\gamma_{\mu}
{\bf{\vec \tau}}/2$, 
we can get the simple relation
\begin{eqnarray}
\label{onshell}
\frac{m_{\pi}^2f_{\pi}}{g_{\pi qq}} = \frac{m_0}{2G}.
\end{eqnarray}
In the chiral limit, $f_{\pi}$ satisfies the Goldberger-Treiman 
relation $f_{\pi}g_{\pi qq}=m$ \cite{Ann}.

\section{Meson corrections at finite temperature and density}

We now extend the above formulae of SU(2) NJL model beyond mean-field approximation 
to finite temperature and density in the  frame of imaginary time temperature
field theory.

At finite temperature using the Matsubara formalism and associated finite 
temperature Feyman rules, one arrives at the same integrals shown in the 
last section with the replacement
\begin{equation}
\int\frac{d^4 p}{(2\pi)^4} \rightarrow 
\frac{i}{\beta}\sum_n\int_0^{\Lambda_F}\frac{d^3 {\bf\vec p}}{(2\pi)^3}
\end{equation}
for quark integration, with the zero-component of momentum $p_0$ 
replaced by $i\omega_n+\mu=(2n+1)i \pi T+\mu$, and
\begin{equation}
\int\frac{d^4 q}{(2\pi)^4} \rightarrow 
\frac{i}{\beta}\sum_n\int_0^{\Lambda_M}\frac{d^3 {\bf\vec q}}{(2\pi)^3}
\end{equation}
for internal meson integration, with the zero-component of momentum $q_0$
replaced by $i\nu_n=2n\pi i T$.
Here $T$ is temperature, $\mu$ the chemical potential,
and the sums on $n$ run over the Matsubara
frequencies $\omega_n$  for quarks and $\nu_n$
for mesons. 

While the $T=0$ 
calculation could be performed using covariant, as well as 
non-covariant regularization 
prescriptions, at 
$T\not= 0$ Lorentz invariance is always broken and the physical $O(3)$ regularization
restriction $|{\vec p}|<\Lambda$ presents itself most naturally. We take the pragmatic
approach of applying this at finite temperature, and consider the device of the
regularization scheme as part of the definition of the model. After the summation 
over the frequencies for quarks and internal mesons, one takes for external mesons 
the extension
$i\alpha_l \rightarrow k_0=m_M$.  

Evaluation of frequency summation in two- and three-loop diagrams is not trivial, 
especially when two frequencies appear in a quark propagator or a meson 
propagator
resulted from the energy conservation in Feynman rule.

\section{Numerical results at finite density}

As discussed in the beginning of this paper, the quark condensate behaves quite
differently at finite density compared with that at finite temperature.
The meson effects on chiral condensate at finite temperature has been investigated
in \cite{mesonT1} and \cite{mesonT2}.  In this paper,
we focus our attention on the meson effects at finite density.
The second reason to choose numerical study at finite density is technical:
At $T=0$ any frequency sum in imaginary time formulation 
is reduced to a step function of chemical potential which
simplifies the momentum integration greatly.

As mentioned above, we introduced in our calculations the momentum
cut-off $\Lambda_F$ for quarks, $\Lambda_{M}$ for pions and
sigmas.  We can not 
determine the four parameters for SU(2) NJL model, $m_0$,
$G$, and the three momentum cut-off $\Lambda_F$ and $\Lambda_{M}$,
since we know  only two experimental observables
$m_{\pi}=139 MeV$ and $f_{\pi}=93.2 MeV$ and an empirical value of 
quark condensate $<{\bar q}q>=-(250 \pm 50 {\rm MeV})^3$.

In \cite{mei2}, we have numerically investigated
the NJL model parameters by fitting
pion mass and pion decay constant with an appropriate 
current quark mass $m_0=5.5MeV$. The model parameters in the mean field approximation are
$\Lambda_F=637.7MeV$, $G \Lambda_F^2=2.16$,  
and $-<{\bar q} q>^{1/3}=248.0 MeV$ with quark
mass $m=330 MeV$. Going beyond mean-field approximation, we choose two series 
of NJL parameters: $\Lambda_F=562.3MeV$, $G \Lambda_F^2=2.31$, and 
$-<{\bar q} q>^{1/3}=223.2 MeV$ with $\Lambda_M/m=1.5$ and $m=330 MeV$,  
and $\Lambda_F=515.7MeV$, $G \Lambda_F^2=2.34$, $-<{\bar q} q>^{1/3}=218.2 MeV$ 
with $\Lambda_M/m=2$ and $m=370 MeV$. The degree of meson fluctuations is controled 
by the ratio $\Lambda_M/m$.
It was found in \cite{mei2}, that for the first group parameters, the meson correction on 
chiral condensate is about $30 \%$ ,  and for the second group parameters, the meson correction
is about $40 \%$. The big correction in the second case arises from the big meson fluctuations
characterized by the parameter $\Lambda_M/m$.

\begin{figure}[ht]
\vspace*{-4truecm}
\centerline{\epsfxsize=15cm\epsffile{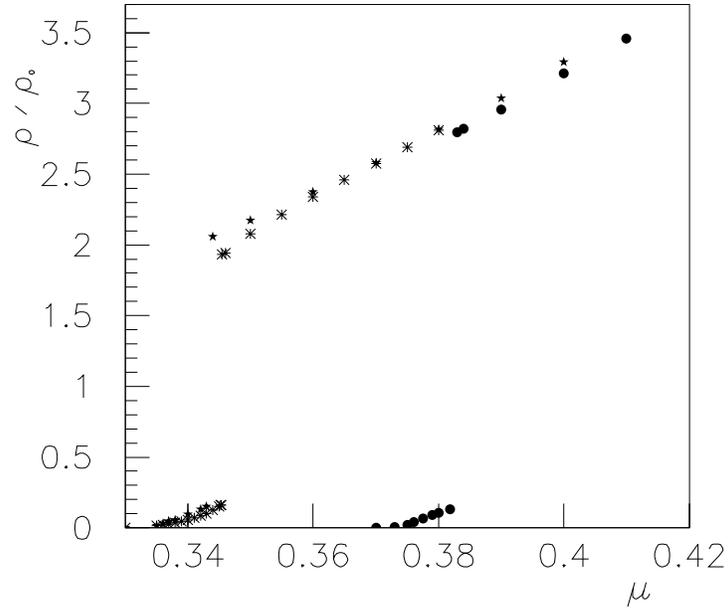}}
 
\caption
{The scaled baryon density as a function of chemical potential $\mu$,   the stars 
correspond to the results in mean-field approximation, the pentagrams and the
solid circles correspond to the results beyond mean-field approximation with the 
first and second group parameters, respectively.  }
\label{rhomu_fig}
\end{figure}

Our results of investigating chiral phase transition by using the above parameters 
will be shown as a function of the scaled density 
$n_b=\rho / \rho_0$, where $\rho_0=0.17 fm^{-3}$ is the normal nuclear matter density,
and  $\rho=\frac{1}{3} < \psi^{\dagger} \psi>=\frac{1}{3} < {\bar \psi} \gamma^0 \psi>$ 
is the baryon density of the system,  which has been calculated to the next-to-leading order 
in $1/N_c$ expansion. The scaled baryon density as a function of
chemical potential $\mu$ is shown in 
Fig. \ref{rhomu_fig},   the stars correspond to the results in mean-field 
approximation, the pentagrams and the solid circles correspond to the results 
beyond mean-field approximation with the first and second group 
parameters, respectively.  

The jumps in Fig. \ref{rhomu_fig} show that chiral phase transition at finite density is of first 
order both in and beyond mean-field approximation. 
For the first group parameter,  the lower critical density $n_b= 0.15$ is a little bit smaller 
than that in the mean-field approximation $n_b=0.162$, and the upper
critical density $n_b=2.1$ is a little bit larger than 
the mean-field value $n_b=1.93$. For the second group parameter,
 it is found that the lower critical density $n_b= 0.13$ is $20 \%$  smaller than the
 mean-field value, and the
upper critical density $n_b=2.79$ is $40 \%$  larger than 
that in mean-field approximation. 
In both cases, the meson corrections 
speed up the  chiral phase transition, which is similar to the results 
at zero density and  finite temperature in  \cite{mesonT1}, and extend the mixed phase of 
the first-order chiral transtion, while the degree of the change depends on the
magnitude of the meson fluctuations.

\begin{figure}[ht]
\vspace*{-4truecm}
\centerline{\epsfxsize=13.5cm\epsffile{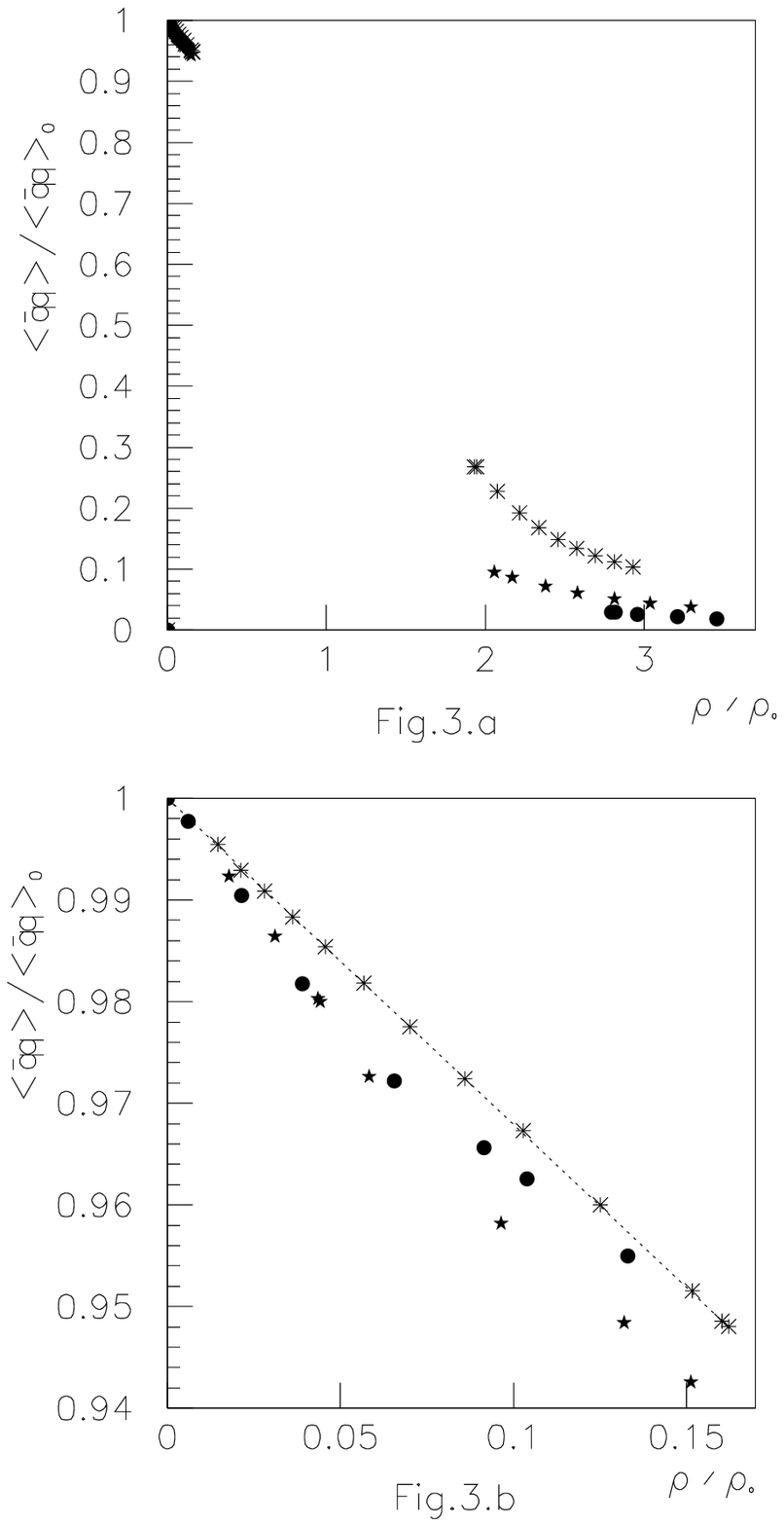}}
 
\caption
{The scaled quark condensate in the whole density region (a) and only before the 
chiral phase transition (b) as a function of the 
scaled baryon density. The stars correspond to the results in mean-field 
approximation, and the pentagrams and the solid circles correspond to the results 
beyond mean-field approximation with the first and second group of parameters.
The dashed line is used to represent the linear relation in mean-field case.}
\label{ratio_fig}
\end{figure}

We show the quark condensate
scaled by its vacuum value $<{\bar q} q> / <{\bar q}q>_0$ in Fig. \ref{ratio_fig}. $a$,
as a function of the scaled density $n_b=\rho / \rho_0$.
It is found that the quark condensate decreases to $30 \%$ of its vacuum value in the mean-field
approximation, while it decreases to $10 \%$ of the vacuum value when 
meson corrections are included for the first group parameters, 
and even decreases to nearly zero for the second group parameters. This means that the meson
corrections can help  melt the chiral condensate more completely.  

To show the meson effects at low density more clearly, we  pick out the low density part of 
Fig. \ref{ratio_fig}. $a$  and enlarge it in  Fig. \ref{ratio_fig}. $b$. It is found that 
in the mean-field approximation, the chiral condensate
and baryon density satisfy the linear relation \cite{rho}
\begin{equation}
\frac{<{\bar q} q>}{<{\bar q} q>_0}=
1-\frac{\Sigma_{\pi N}}{f_{\pi}^2m_{\pi}^2}\rho
\end{equation}
before the phase transition,
where the nucleon sigma term $\Sigma_{\pi N}=0.043 {\rm GeV}$
is in good agreement with the experimental value  
$0.045 {\rm GeV}$  extrapolated
from low energy pion-nucleon scatterings. 
However, when meson corrections are considered, this linear relation is more or less broken,
depending on the degree of meson fluctutions.

\section{Conclusions}

In this paper we extended the self-consistent scheme of SU(2) NJL model 
beyond mean-field approximation and with chiral symmetry explicitly breaking 
to finite temperature and density, and calculated numericaly 
the meson corrections to the
quark condensate at finite density.  

The numerical results show that  the meson fluctuations can  
significantly change the properties of the system at finite density. Compared with the results in 
mean-field approximation, 
the meson fluctuations 1) speed up the chiral phase transition, 
2)  extend remarkbaly the mixed phase of the first-order chiral transition,
3) help melt the chiral condensate more completely, 
4) break the linear relation between the quark condensate and the baryon 
density.

\section*{Acknowledgements}

    This work was supported in part by the NSFC under Grant No. 19925519 and the Major State Basic
Research Development Program under Contract No. G2000077407.

\newpage

\newpage

\newpage

\end{document}